# Unveiling the complex structure-property correlation of defects in 2D materials based on high throughput datasets

Pengru Huang[1], Ruslan Iuckin[2], Maxim Faleev[2], Nikita Kazeev[1,3], Abdalaziz Rashid Al-Maeeni[4], Daria V. Andreeva[1,5], Andrey Ustyuzhanin[3,4], Alexander Tormasov[2], A. H. Castro Neto[1,5], Kostya S. Novoselov[1,5*]

[1] Institute for Functional Intelligent Materials, National University of Singapore, 117544, Singapore
[2] Innopolis University, Universitetskaya St 1, Innopolis, Republic of Tatarstan, Russia, 420500
[3] Schaffhausen Institute of Technology, Rheinweg 9, 8200 Schaffhausen, Switzerland
[4] HSE University, Myasnitskaya Ulitsa, 20, Moscow, Russia, 101000
[5] Centre for Advanced 2D Materials, National University of Singapore, 117546 Singapore
*kostya@nus.edu.sg

## Abstract

Modification of physical properties of materials and design of materials with on-demand characteristics is at the heart of modern technology. Rare application relies on pure materials – most devices and technologies require careful design of materials properties through alloying, creating heterostructures of composites or controllable introduction of defects. At the same time, such designer materials are notoriously difficult for modelling. Thus, it is very tempting to apply machine learning methods for such systems. Unfortunately, there is only a handful of machine learning-friendly material databases available these days. We develop a platform for easy implementation of machine learning techniques to materials design and populate it with datasets on pristine and defected materials. Here we describe datasets of defects in represented 2D materials such as $MoS_2$, $WSe_2$, hBN, GaSe, InSe, and black phosphorous, calculated using DFT. Our study provides a data-driven physical understanding of complex behaviors of defect properties in 2D materials, holding promise for a guide to the development of efficient machine learning models. In addition, with the increasing enrollment of datasets, our database could provide a platform for designing of materials with predetermined properties.

**Introduction**

Intelligent design of materials with predetermined properties is the central agenda for modern materials science. To this end multiple materials genome projects have been established. Two-dimensional (2D) materials is a very important subset of our overall material library, and, thus, it is not surprising, that many genome projects have been associated exactly with 2D crystals. Still one of the most exciting opportunities provided by the 2D materials is the possibility to controllably alter their properties through chemical modification – materials without bulk are strongly susceptible to the introduction of foreign atoms either at the surface as adatoms or into the plane of the crystal as substitution. Still, up to now, no database is available which would deal with the modification of the properties of 2D crystals by the introduction of defects. Here we present our database for 2D materials and defects in such crystals, which offers comprehensive analysis of such structures with the use of machine learning algorithms (will be described elsewhere) and to analyze the electronic properties of defects in $MoS_2$ and $WSe_2$.

Defects play a very important role in terms of modification of the mechanical, thermal, electronic, optical and other properties of solids. Thus, individual defects can act as single photon emitters, qubits, be used for single atom catalysis and further used for many other applications[1]. Individual defects have been demonstrated to be even more important for the modification of properties of 2D materials[2,3]. For instance, it has been shown that localized defects embed in transition metal dichalcogenides (TMDs) and hexagonal boron nitride (h-BN) exhibit single-photon emission (SPE) at cryogenic and room temperatures, which hold advantages over the 3D counterpart of NV centers in diamond[4,5].

Controllable defect engineering for the purpose of the modification of material's properties requires knowledge of the structure-property relation of defects, which is a formidable task considering the vast space of possible host materials and defect configurations. Even though a single defect structure can be calculated with state-of-the-art density functional theory (DFT) methods in modern high performance computing infrastructures within hours, such computations are not scalable. The properties of each new defect has to be calculated from scratch. The state-of-the-art research paradigm integrated of high throughput simulations, data science, and machine learning is appealing to accelerate material exploration. In this spirit, a large number of computational databases such as Material Projects, the Open Quantum Materials Database (OQMD), the Automatic Flow for Materials Discovery (AFLOWLIB), and the Novel Materials Discovery (NOMAD) Laboratory have been established[6–9]. Various machine learning (ML), especially graph neural networks based methods for material science such as MEGNet, CGNN, SchNet, GemNet, etc. have been proposed for property prediction[10–13]. The increasing repositories of computational datasets and the development of machine learning methods together have led to an exploding growing of in-silico material exploration in the areas of 2D materials, catalyst, batteries, and photovoltaics, etc[14–18].

Still, great difficulties arise when trying to apply machine learning to predict properties of defects, which may be due to the lack of defect datasets and the challenge in the prediction of quantum states. Also, there is a great deal of uncertainty when the machine learning of "black box" nature encounter the nonlinear quantum properties in defects. Due to these reasons, there have been few studies of machine learning applied to defects in solids. The reported studies mainly focus on the prediction of the key properties of point defects in 2D materials, predictions of vacancy migration and formation energies in alloys, and defect dynamics in 2D TMDCs[19–21]. Even though great efforts have been devoted to the screening and creation of database of 2D materials[15,16,22], such effort on defect properties is limited. Only recently, Fabian et. al. reported a quantum point defect database in a size of 503 defect structures in 82 different 2D materials[23]. The study is comprehensive and including a wide range of thermodynamic and electronic properties. However, according to our ongoing exploration of machine learning defect prediction, the size and the density of data is still limited, thus hindering efficient prediction with AI methods. A comprehensive dataset of defect properties is necessary not only for the engineering of imperfections in solids towards emerging technologies, but also for the perfection of machine learning models appropriate to defect prediction.

In this paper, we present datasets of defect properties in represented 2D materials created by high throughput DFT calculations. The property distribution in the dataset can be visualized through a binary property map. Interestingly, the band gap vs. formation energy plot show a non-trivial correlation which is attributed to the hierarchical impact of defect components on the host crystal. Such property maps could be seen as fingerprints of 2D crystals and defects introduced. The interaction between defects decay with distance in the oscillatory manner due to the contributions of quantum phenomena. Such quantum oscillations are understood via a simplified quantum mechanics model and visualized by the wave function resonance of defect states. Our study provides a data-driven physical understanding of the electronic properties of defects in 2D materials, which would help better design of machine learning models towards accurate and efficient prediction. Moreover, our datasets could serve as a platform for competitive model training, and with the enrollments of more and more datasets, scalable defect prediction would be expected.

**Methods**

We computed datasets of both simple defects and high density defects in 2D materials. For simple defects, all possible symmetrically inequivalent single, double, and triple site defects composed of Mo vacancies, S vacancies, W substitutions, and Se substitutions in the 8x8 monolayer $MoS_2$ supercell were created. The compositions, configuration numbers, atomic structures of each type of defects are summarized in Table1. The same procedure was repeated for $WSe_2$ with substitution atoms of Mo and S instead. The dataset, containing DFT computational properties of 5933 defect configurations for MoS$_2$ and another 5933 for WSe$_2$, is available online[23] (https://ro-los.com/open/2d-materials-point-defects/). Another dataset (also available in the same repository) of high density defects was created by randomly generating combined vacancy and substitution defects in a wide range of concentration of 2.5%, 5%, 7.5%, 10%, and 12.5% for represented 2D materials such as MoS$_2$, WSe$_2$, hBN, GaSe, InSe, and black phosphorous (BP). We generated and computed 100 structures for each defect concentration for each material, totaling 500 configurations for each material and 3000 in total. Availability of such high-density defects datasets, alongside with the full set of triple site defects will allow researchers to verify complex scenarios which might arise in defect engineering.

Our calculations are based on density functional theory (DFT) using the PBE functional as implemented in the Vienna Ab Initio Simulation Package (VASP)[25–27]. The interaction between the valence electrons and ionic cores is described within the projector augmented (PAW) approach with a plane-wave energy cutoff of 500 eV[28]. Spin polarization was included for all the calculations. The initial crystal structures were obtained from the Material Project database and the supercell sizes and the computational parameters for each materials are listed in the supplemental materials (Table S1). Since very large supercells are used for the calculation of defect, the Brillouin zone was sampled using Γ-point only Monkhorst-Pack grid for structural relaxation and denser grids for further electronic structure calculations. A vacuum space of at leat 15 Å was used to avoid interaction between neighboring layers. In the structural energy minimization, the atomic coordinates are allowed to relax until the forces on all the atoms are less than 0.01 eV/Å. The energy tolerance is $10^{-6}$ eV. For defect structures with unpaired electrons, we utilize standard collinear spin-polarized calculations with magnetic ions in a high-spin ferromagnetic initialization (the ion moments can of course relax to a low spin state during the ionic and electronic relaxations). Currently, we are focusing on basic properties of defects and did not include spin orbit coupling (SOC) and charged states calculations[29].

The formation energy, i.e., the energy required to create a defect is given by

$$E_f = E_D - E_{pristine} - \sum_i n_i \mu_i \quad (1)$$

where $E_D$ is the total energy of the defect structure, $E_{pristine}$ is the total energy of the pristine MoS$_2$ or WS$_2$, $n_i$ is the number of an element (Mo, W, S, or Se) transferred from the defect structure to the chemical reservoirs, and $\mu_i$ is the chemical potential of the element.

The interaction energy for a defect complex with respective to the defect components is defined as

$$E_{int} = E_f(D) - \sum_i E_f^i \quad (2)$$

Where $E_f(D)$ is the formation energy of the double-site or triple-site defect complex and $E_f^i$ is the formation energy of the single-site defect components. Negative (positive) interaction energy indicates the tendency for the individual defects to attract (repel) each other.

To parameterize the electronic properties of defects, we inspect the positions of the highest occupied states, the lowest unoccupied states, and the separation of these two levels of defects structures. For the sake of representation, we adopt the terminologies of the highest occupied molecule orbital (HOMO), the lowest unoccupied molecule orbital (LUMO), and band gap for these electronic properties. This is reasonable, considering that the defects in TMDCs are highly localized and molecule-like. Practically, we would like to evaluate the positions of defect levels with respect to the valence band maximum (VBM). However, due to the finite cell effect, the calculated valence band edge is generally highly disturbed by defects and hardly identified. Moreover, the whole Kohn-Sham energy spectrum shifts due to the introduction of defects. For these reasons, we chose the deepest Kohn-Sham orbital as a reference with the considering that it is less affected by defects. Accordingly, the HOMO of the defects with respect to the pristine VBM are normalized according to

$$E_{HOMO} = E_{HOMO}^D - E_1^D - (E_{VBM}^{pristine} - E_1^{pristine}) \qquad (3)$$

Where $E_{HOMO}^D$ is the energy of the highest occupied Kohn-Sham state of a defect at the $\Gamma$ point, $E_{VBM}^{pristine}$ is the energy of the valence band maximum of pristine MoS$_2$ or WS$_2$, $E_1^D$ and $E_1^{pristine}$ are the energy of the first Kohn-Sham orbital of the calculated defect and pristine MoS$_2$ or WS$_2$ structures. Since the defect states are localized – the defect bands are flat in the wave vector space. Thus, the variations in energy in the k-space are small, and we extracted all the energies at the $\Gamma$ point. The regularized LUMO energy is defined similarly as

$$E_{LUMO} = E_{LUMO}^D - E_1^D - (E_{VBM}^{pristine} - E_1^{pristine}) \qquad (4)$$

There is well known underestimation of band gap at the level of PBE functionals. To well reproduce the experimental band gap, more advanced hybrid functional or many-body interaction including methods should be employed. However, such methods are computationally much more expensive and impractical for high throughput calculations. Since we are focusing on the general trends of formation energy, HOMO, and LUMO for a wide range of defect structures, the main conclusions based on PBE functionals should be transferable to results of other methods.

**Results and Discussion**

The defect space in 2D materials is mainly attributed to three aspects of variables, i.e. the 2D material hosts, the defect components, and the defect configurations (**Figure 1a**). There are hundreds of existing 2D materials and even more are expected to be synthesized in the future. The components of defects are limited to a few types such as intrinsic vacancies, impurity substitu-tions, and antisites. However, the configuration space for defects is infinite. These three variables together contribute a vast defect space for 2D materials which is impractical for thorough experimental or even computational investigations.

Considering the extremely large space of defects, probably any practical datasets should be termed as "small data". Consequently, to train machine learning methods for efficient prediction of defect properties, the structure of the datasets should be carefully designed. Our datasets are established in two groups, including one of structured and the other of dispersive (**Figure 1b**). The structured datasets contain fine-tune features of defects correlated with the periodic nature of crystal lattices and the quantum mechanics nature of defect properties. Herein, as one of example, we generate and compute a structured dataset of TMDCs by screening all possible single, double, and triple site defects in $MoS_2$ and $WSe_2$. The dataset include properties of simple defects of different configurations and inter-defect separations, and provide very detail information for machine learning. This dataset contains dense data and occupies only a small subspace of the whole defect space. The other group of dispersive datasets aim to spread data in the large defect space. This is done here by calculating the properties of a wide concentration range of defects for a wide range of 2D materials. So far, we have finished defect properties of a few popular materials including $MoS_2$, $WeSe_2$, hBN, GaSe, InSe, and black phosphorus. Such random sampling strategy has a disadvantage in the sparsity of data. However, it allows us to cover the diverse phase space of defect properties and facilitates design of universal, transferable machine learning algorithms such as active learning[30].

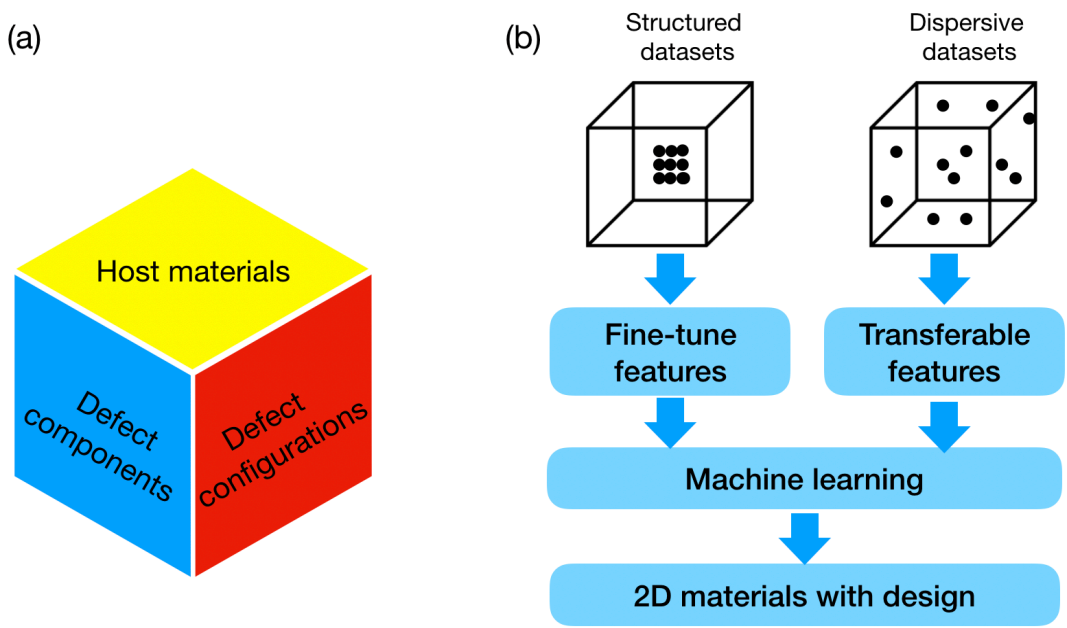

**Figure1** (a) The space of defects in 2D materials and (b) the structure of the datasets for machine learning of defect properties.

# 1. A summary of the datasets

The structured dataset is composed of DFT-computed properties of 5933 defect configurations for $MoS_2$ and another 5933 configurations for $WS_2$. The compositions, numbers, the atomic structures, and the variation range of properties of each type of defects for $MoS_2$ are summarized in **Table 1**. The dispersive dataset includes computational properties of defects in a wide range of concentrations for representative 2D materials such as $MoS_2$, $WSe_2$, hBN, GaSe, InSe, and black phosphorus. The datasets together with the output data such as relaxed atomic structures, density of states (DOS), and band structures in the DFT PBE levels are available for further studies and machine learning training[23].

**Table 1** The composition, number, structure, and the variation range of the formation energy, HOMO, and LUMO for each type of defects created in the 8x8 supercell of $MoS_2$. The color code for atomic structures has Mo in purple, S in yellow, W in red, Se in green, and S vacancy in white. The symbol X, V, and S represent defect types of (1) S sites only, (2) one Mo vacancy plus S sites, and (3) one W substitution plus S sites.

| Type | Defect site | | | Count | Variation range | | | Structure |
|---|---|---|---|---|---|---|---|---|
| | Mo | S | S | | Energy (eV) | HOMO (meV) | LUMO (meV) | |
| X1 | | vac | | 1 | | | | 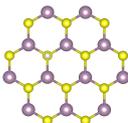 |
| X2 | | vac | vac | 19 | 0.113 | 105 | 358 | 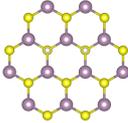 |
| X3 | | Se | | 1 | | | | 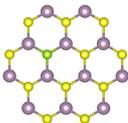 |
| X4 | | Se | Se | 19 | 0.012 | <10 | <10 | 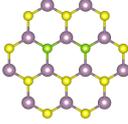 |

| | | | | | | | |
|---|---|---|---|---|---|---|---|
| X5 | | vac | Se | 29 | 0.106 | <10 | <10 | 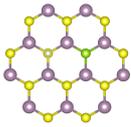 |
| V1 | vac | | | 1 | | | | 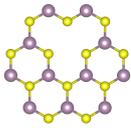 |
| V2 | vac | vac | | 15 | 1.992 | 59 | 124 | 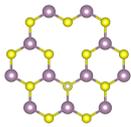 |
| V3 | vac | vac | vac | 743 | 4.121 | 328 | 450 | 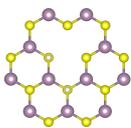 |
| V4 | vac | Se | | 15 | 0.197 | 129 | 25 | 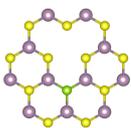 |
| V5 | vac | Se | Se | 743 | 0.430 | 245 | 73 | 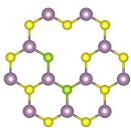 |
| V6 | vac | vac | Se | 1415 | 2.282 | 203 | 236 | 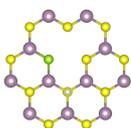 |
| S1 | W | | | 1 | | | | 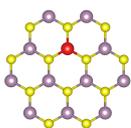 |
| S2 | W | vac | | 15 | 0.042 | 15 | 24 | 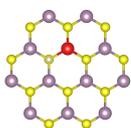 |
| S3 | W | vac | vac | 743 | 0.193 | 108 | 394 | 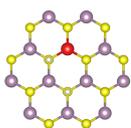 |
| S4 | W | Se | | 15 | 0.028 | <10 | <10 | 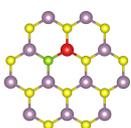 |
| S5 | W | Se | Se | 743 | 0.079 | <10 | <10 | 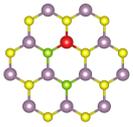 |
| S6 | W | vac | Se | 1415 | 0.167 | <10 | <10 | 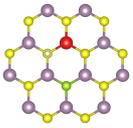 |

The variation range in the formation energy provides knowledge about the amplitude of the interaction between the defects for each defect type. The formation energy for V3 defects composed of one Mo vacancy and two S vacancies span the widest range of ~4.1 eV, whereas defects that included one Mo vacancy and only one S vacancies varies within a moderate range of ~2.2 eV. The variation in the formation energy of defects contain two S vacancies is only 0.1 eV. The variation of the formation energy of defects composed mainly by substitution atoms is only a few tens of meV.

Such a spread of energies for different types of defects seems logical. The creation of a Mo vacancy breaks six Mo-S bonds and induces the most significant lattice imperfection. A large formation energy of Mo vacancy of 7.12 eV is needed to be paid for such imperfection. A creation of S vacancy requires breaking of only three Mo-S bonds and thus produces a moderate disturbance to the lattice and requires a formation energy of 2.65 eV. Due to the similar outer electron configuration between W and Mo, and between Se and S, substitution atoms have almost negligible change in the house lattice and electronic structures. The formation energies for W substituted Mo, and Se substituted S, are 0.167 eV and 0.279 eV respectively. The impurity orbitals of both of these two substitutions are merged into the valence band or conduction band of $MoS_2$ and do not show any defect level in the band gap.

The variation ranges of the frontier defect orbitals (HOMO and LUMO) for each type of defect provide knowledge about to what extend we can manipulate the defects to achieve desired properties. This information is of great importance for such applications as quantum computing and quantum telecommunications (single photon emitters), catalysis and many others. Vacancy defects create deep-lying levels that span a range of 0.1-0.4 eV inside the band gap. At the same time, substitutional defects create defect states inside the valence or conduction bands without significantly change to the band edges – of the order of 10 meV.

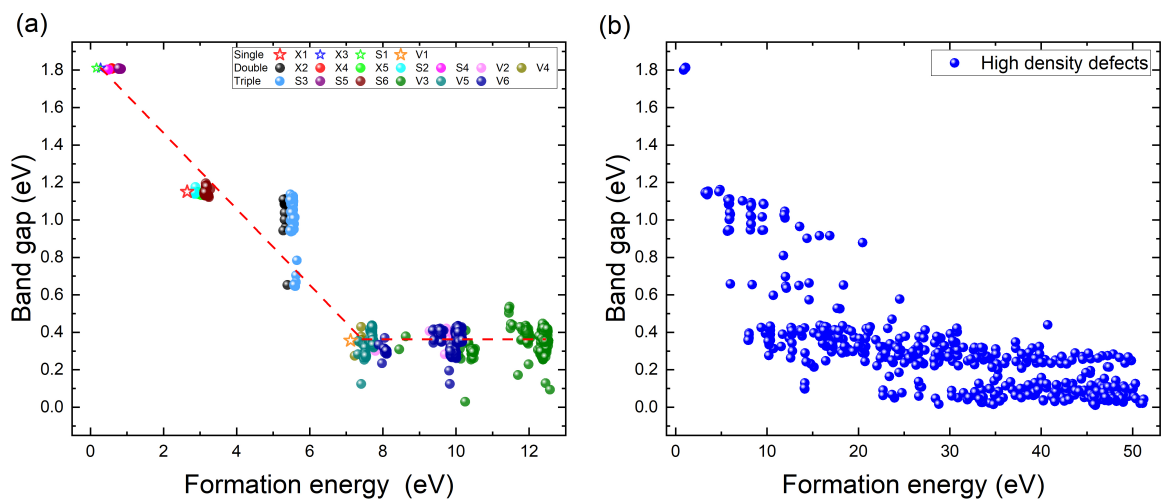

**Figure 2** The binary property (band gap vs. formation energy) map of defects in MoS$_2$. The property maps for (a) simple defects and (b) high density defects. The band gap is calculated as the separation between HOMO and LUMO. The data for each defect type are color coded and the dash line is a guide for the eyes in (a).

To visualize the distribution of the whole dataset in the property space, we show the binary property, i.e. band gap vs formation energy plot in **Figure 2**. Most interestingly, as shown by the guide line in **Figure 2a**, there is an overall trend for the whole dataset that the band gap of defect structures decreases with increasing formation energy and finally converge in groups near 0.3 eV. In other words, to achieve the deep defects levels inside the band gap (which reduces the gap) – a large formation energy has to be paid. This provides a guide line for the manipulation of defects in TMDCs in order to achieve the desired properties. Thus, W and Se substitutions produce defect levels without a significant alteration of the electronic structure. On the other hand, shallow gap states can be introduced by single or double S vacancies. Formation of deep levels requires the presence of Mo vacancy.

The distribution of defects in the binary property plot can be understood by the hierarchy impact of defects to the host MoS$_2$. For the substitution only defects (X3, X4, S1, S4, S5), since the disturbance on the lattice is negligible (< 10 meV) - the formation energies are low. And because no gap state is introduced - the band gap of the substitution defects retains the separation of band edges of MoS$_2$ of 1.81 eV. For defects composed of one S vacancy (X1, S2, S6), the band gaps concentrate at ~1.1 eV and the formation energies lie near ~3.0 eV. For defects composed of two vacancies (X2, S3), the formation energy is approximately double of that of single S vacancy ~5.5 eV. Interestingly, even though the formation energy has a very low span, the band gaps of X2 and S3 span in a wide range of 0.6 eV. This indicates that, by controlling the distance between S vacancies, the frontier orbitals of defects become highly tunable. For defects that include Mo sites (V1, V2, V3, V4, V5, V6), the band gaps are mainly within 0.5 eV., The formation energy span over 7.0-13.0 eV, depending on the number of vacancies.

As shown in **Figure 2b**, the trend in the property map of simple defects persists for the high density defects of $MoS_2$. Considering that in the high density defect dataset, there are wide spread of the configurations and concentrations of defects, the non-trivial distribution of properties is intrinsic and could be serve as a fingerprint for defects in 2D materials. Indeed, the property distribution maps for $MoS_2$, $WSe_2$, hBN, GaSe, InSe, and black phosphorus shown distinct characteristics for different 2D material types. That is, the maps for $MoS_2$ and $WSe_2$, or GaSe and InSe share the similar features, while largely different between different crystals. This is shown in the supplemental materials (**Figure S3**).

Defects with unpaired electrons are of special interest for magneto-optical and information technology applications. Among our datasets, no magnetic defects are found in $MoS_2$ and $WSe_2$ while such defects could be created in GaSe, InSe, BP, and C-doped hBN. The exchange splitting of states results in asymmetric distribution of band gap for majority and minority states. Interestingly, there seems to exist a nontrivial trend of larger energy distribution of majority band gap. This is especially conclusive for defects in BP and C-doped hBN. Vacancy defects in GaSe, InSe, and BP are intrinsically possess local magnetic moments due to the odd number of valence electrons of Ga, In, and P atoms. The introduction of a C impurity at both the B and N sites create an unpaired electron due to the valence electron difference. The total spin of the carbon-impurity complexes in hBN is govern by the difference between the difference between the carbon substitutions at the two sublattices according to Lieb's theorem[31,32].

$$S = \frac{N_{C(N)} - N_{C(B)}}{2} \tag{5}$$

For this reason, the magnetic moments in C-doped hBN could reach a very large number if the carbon substitutions in the two sublattices are highly imbalanced (**Figure S4**). Carbon substitutions in hBN are particularly interesting for the area of single photon emitters aiming at the identification and engineering of candidate emitting centers in 2D hosts instead of that in bulk materials such as the NV center in diamond[32,33].

**2. The quantum fluctuation of defect properties**

Complex defects in TMDCs demonstrate non-trivial quantum properties. Thus, for a V2 defects (one Mo vacancy and one S vacancy) the interaction energy (calculated as the formation energy difference between the defect complex and the sum of the defect components), as well as HOMO and LUMO, exhibit oscillatory behaviour as a function of the separation between the vacancies, **Figure 3**. The oscillatory behaviour is most obvious for the small separation between the vacancies. The local minima in the interaction energy correspond to the configurations when the S vacancies are at the 1st, 3rd, 6th and 10th nearest S sites to the Mo vacancy. These are exactly the triangular number series $S_n = n \cdot (n+1)/2$. Interestingly, the sites which exhibit local minima of the formation energy lie exactly along the zigzag crystalline orientation, as shown in the inset of **Figure 3a**.

In metals the introduction of impurities generally results in the Friedel oscillations[34]. Due to the strong screening of the surrounding electrons, the long-range tail of a charged impurity potential is suppressed, resulting in a power law decay of the defect disturbance[35,36]. In insulators, however, when the wave-function is strongly localized on the defect site - the lattice structure plays a significant role on the defect properties[1,37,38].

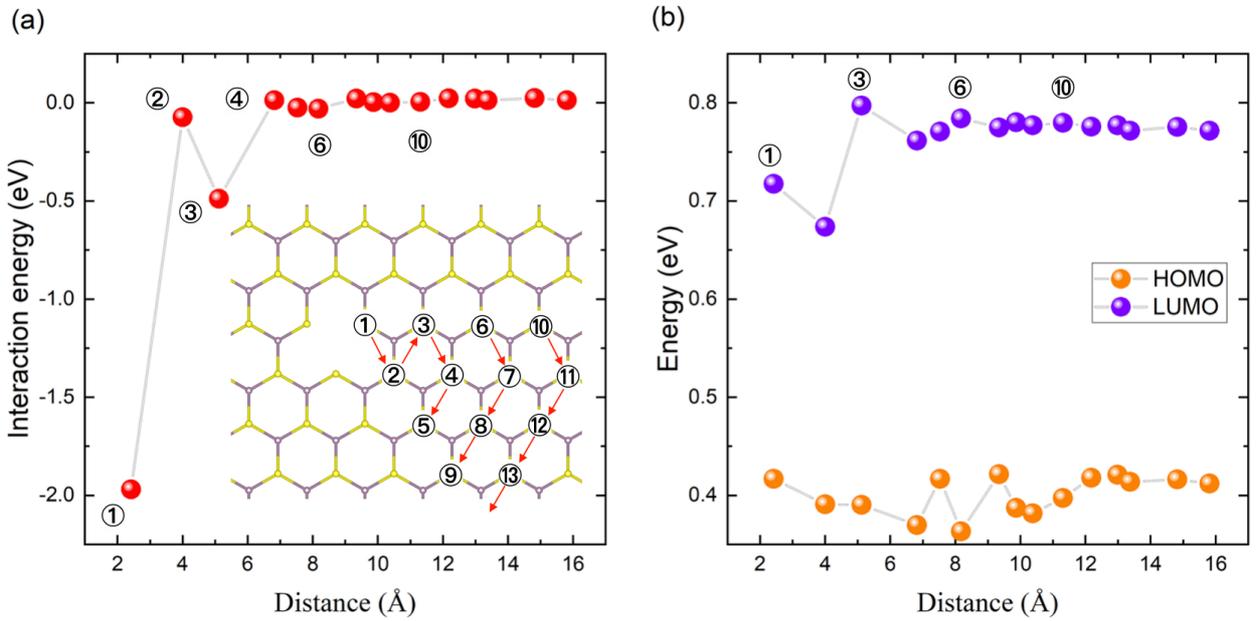

**Figure 3** The (**a**) interaction energy, (**b**) HOMO, and LUMO for V2 defects as a function of distance between the Mo vacancy and the S vacancy. The inset labels the positions of the nearest S sites to the Mo vacancy.

In crystals with honeycomb structure the interaction between defects is strongly influenced by the presence of the two sublattices. Thus, it has been predicted and confirmed both by theory and experiment that the coupling between hydrogen atoms on the surfaces of graphene depends on the sublattices occupied[39,40]. To understand the defect interaction in TMDCs, we employ a simplified two-orbital picture. We propose that there are two localized defect orbitals $\phi_A$ and $\phi_B$ at the lattice site A and B of the crystal, each was occupied by one electron. In case that the two defects are largely separated, the correlation between the two is negligible. Each single defect wave function fulfill the single particle Schrödinger equation

$$(T_1 - V_A^1)\phi_A = E_1\phi_A \tag{6}$$

$$(T_2 - V_B^2)\phi_A = E_2\phi_B \tag{7}$$

where $T$ is the kinetic energy of the electron, $V_A^1$ and $V_B^2$ represent the Coulomb potential energy at the A and B sites contributed together by the atomic nuclei and band electrons. As the two defects occupied neighboring lattice sites, the coupling of the two should be taken into account, and the two-electron wave function takes form:

$$\Phi(r_1, r_2) = \frac{1}{\sqrt{2(1 \pm S^2)}}[\phi_A(r_1)\phi_B(r_2) \pm \phi_A(r_2)\phi_B(r_1)] \tag{8}$$

Where $S = \langle \phi_A | \phi_B \rangle$ is the overlap integral of the two orbitals. The positive and negative signs correspond to the bonding and anti-bonding states, respectively. The Hamiltonian of the defect complex is:

$$H = (T_1 + T_2 - V_A^1 - V_B^2) - V_B^1 - V_A^2 + \frac{e^2}{4\pi\varepsilon_0\varepsilon \cdot r_{12}} + \frac{e^2}{4\pi\varepsilon_0\varepsilon \cdot R_{AB}} \tag{9}$$

In addition to the additive single-defect terms in the parentheses, the Hamiltonian includes the Coulomb potential originating from the neighboring defect sites (represented by $V_A^1$ and $V_B^2$) as well as electron-electron and core-core electrostatic interactions in the screening environment of band electrons (described by the dielectric constant $\varepsilon$). In the last two terms, $r_{12}$ is the distance of the two electrons and $R_{AB}$ is the separation of the two defect sites. The energy of the bonded ($E_+$) and anti-bonded ($E_-$) state can be expressed as:

$$E_\pm = \left\langle \Phi(r_1, r_2) | H | \Phi(r_1, r_2) \right\rangle = \frac{E_1 + E_2}{2} + \frac{K \pm J}{\sqrt{1 \pm S^2}} \tag{10}$$

Where $K$ and $J$ is the direct Coulomb and exchange integrals, which take into account the electron-core attraction, electron-electron repulsion and core-core repulsion of the two defects[32]. The stabilization of the configuration is governed by the competition between the exchange and the direct Coulomb interactions of the two-electron state. According to the definition in equation (2), the interaction energy is essentially the stabilization energy, originating mainly from the exchange interaction of the two defect orbitals:

$$E_{int} = \frac{2(K + J)}{\sqrt{1 + S^2}} \tag{11}$$

Based on the above picture, we can understand the oscillations in the interaction energy of **Figure 3** in detail. Firstly, we check the oscillation of the defect wave function and the displacement of the electric charge of surrounding atoms disturbed by the defect states. A Mo vacancy intro-duces six dangling bonds which result in several energy levels inside the gap (**Figure 4a**). These states inside the gap are highly localized at the vacancy center and decay within a few lattice spaces. As shown in **Figure 4b**, due to the resonance of the electron wave and the honeycomb lattice, the wavefunction of Mo vacancy has nodes at the Mo sites, where it changes sign[36]. This means that this wavefunction demonstrate oscillatory behavior with maxima around S sites and zeros at Mo sites (**Figure 4b**). Likewise, the wave function of a S vacancy demonstrates similar oscillatory behavior, just with the maxima of the wavefunction at Mo sites (**Figure 4c**). This trend is consistent with the fluctuation in the interaction energy of the V2 defects shown in **Figure 3a**. The wave function oscillation of the defect states is reflected in the atomic charges of neighboring atoms. As shown in **Figure 4d**, the atomic charge of S atoms around the Mo vacancy was plotted as a function of distance to the vacancy, showing a similar fluctuation trend in the interaction energy of **Figure 3a**. In a pristine MoS$_2$ structure, the atomic charge of S gained from Mo atoms was calculated to be ~0.6 electron. It is calculated to be less than 0.5 electron for S atoms nearest to the Mo vacancy due to the breaking of the Mo-S bonds. The bond breaking around the vacancy also impacts the atomic charges of the other S atoms due to the wave function fluctuation of defect states. However, such impact decays rapidly due to screening.

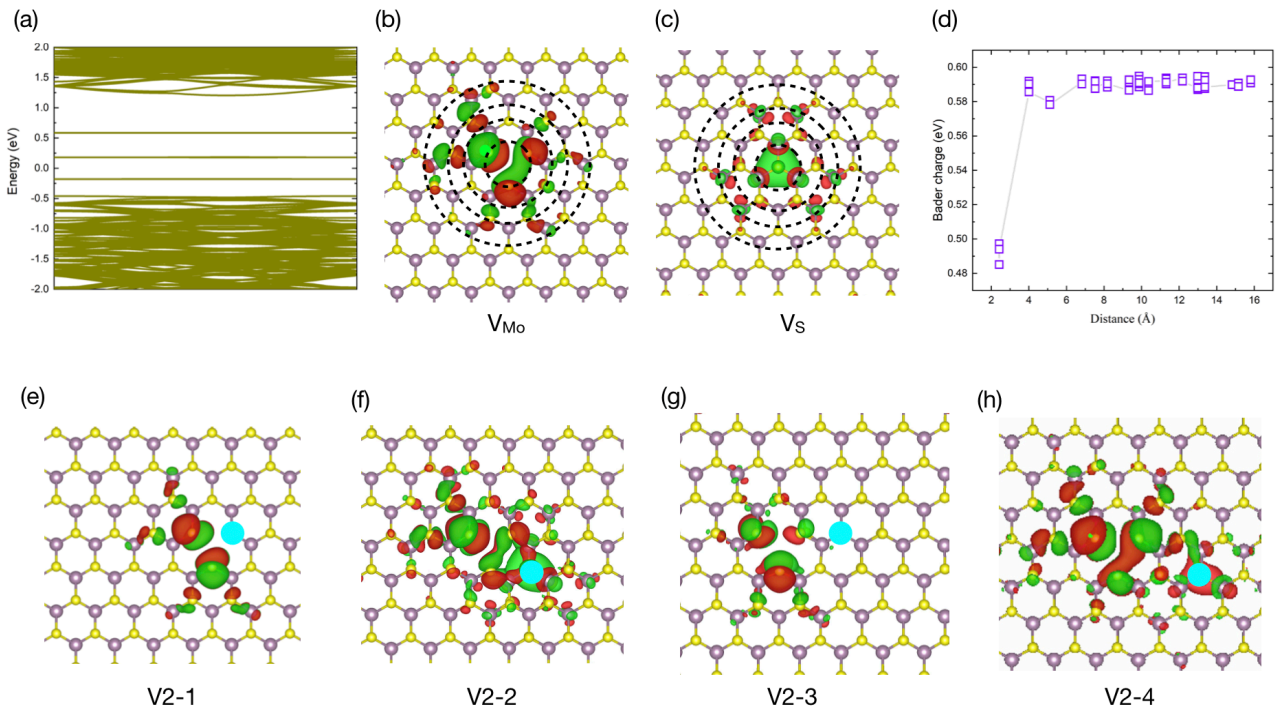

**Figure 4** The orbital coupling of a Mo and a S vacancy in MoS$_2$. (a) The band structures for a Mo vacancy structure. The wave function of HOMO for structures with (b) a Mo vacancy and (c) a S vacancy. The dash circles centering at the vacancies are a guide to the lattice sites. (d) The Bader atomic charge for S atoms around a Mo vacancy in (b). The wave functions for defect configurations with an S vacancy occupy at the (e) 1st, (f) 2nd, (g) 3rd, and (h) 4th nearest site of an Mo vacancy. The cyan dots indicate the S vacancies.

The coupling of the two vacancies is highly site dependent. As the two vacancies occupy neighboring lattice sites, the two states hybridize according to the phase and amplitude of their overlapping wavefunctions. Two states hybridize strongly if their wave functions overlap in-phase, otherwise their coupling is weak (like in the case if their wavefunctions do not overlap or overlap out of phase). This is evident according to the wave functions of the HOMO for the four V2 defects with the S vacancy occupied the 1st, 2nd, 3rd, and 4th nearest lattice site to the Mo vacancy (**Figure 4e-h**). For the 1st and the 3rd configurations, one vacancy occupies the lattice sites where the wave function of the other reaches a peak value, the dangling bond states hybridize strongly which results in a large exchange interaction and stabilization energy. Some of the dangling states of the two vacancies hybridize and lie at the lower energy, leaving other dangling states unaffected. As a result, the HOMO wavefunction is constructed from the unaffected dangling states and shows some features of the pristine Mo vacancy (**Figure 4(e,g))**. For the 2nd configuration, even though the separation between the two vacancies is smaller than that of the 3rd configuration, the two vacancies occupy lattice sites where there is a knot of their wavefunctions. The wavefunctions of the HOMO retain that of the isolated Mo and S vacancies (**Figure 4f**). For defect configurations with the distance larger than the 4th nearest sites, either the separation is too large or the wave function are out of phase - the hybridization can be neglected (**Figure 4g**). According to these wave function plots and the quantum mechanic origin of the stabilization energy, the fluctuation in the interaction energy in **Figure 3** can be understood.

The defect levels in the band gap fluctuate accordingly with the coupling strength of defect states. This can be seen from the opposite variation trend of the LUMO compared with that of interaction energy (**Figure 3b**). The variation in the coupling strength of defect states give rise to diverse locality and affinity of defect electrons, resulting in tunable activity for alien species. On the other hand, a wide range of resonant transitions between defect levels could be achieved owing to different symmetry and separation of defect states. Same rationale can be used to structure the results on the triple defect in which additional complexities involve due to the participation of the third defect sites. These data are presented in Supplementary Information.

**Conclusions**

We develop a database of machine learning friendly datasets on the physical properties of solid state materials with and without defects. As a starting point, a structured dataset of 11866 defect configurations in TMDCs and a dispersive dataset of 3000 configurations in six represented 2D materials were created. It is based on high throughput DFT calculations and unveil the complex structure-property correlations through proper data ordering and physical insights. The initial structured dataset spans over all possible single, double, and triple defects configurations that with components of Mo/W and S/Se vacancies, W/Mo and Se/S substitutions in a 8x8x1 supercell. According to the band gap vs formation energy map, the property distribution of the defect configurations was visualized. This may provide a general guideline for the defect engineering in TMDCs to enable emerging technologies. The property of a selected double and triple defects were studied in detail. The fluctuation of defect properties with the lattice site and distance are observed and explained in depth through symmetry and quantum mechanics electronic structure analysis. The dispersive dataset was created to span over as large as possible in the defect space of 2D materials. Even though the defect configurations in the dataset are highly dispersive, the property maps for the calculated materials persist some non-trial characteristics and exist as fingerprints of the host materials. Our study demonstrates that a properly structured dataset can reveal the complex structure-property correlation. This could provide in-depth guide to the engineering of materials with predetermined properties through their chemical modification, alloying and defect formation. With the enrollment of further datasets of materials with and without defects - a powerful platform for the realization of materials with tailored properties will be formed.

**Acknowledgments**

This research is supported by the Ministry of Education, Singapore, under its Research Centre of Excellence award to the Institute for Functional Intelligent Materials (I-FIM, project No. EDUNC-33-18-279-V12). KSN is grateful to the Royal Society (UK, grant number RSRP\R\190000) for support. PH thanks the supports of the National Key Research and Development Program (No. 2021YFB3802400) and the National Natural Science Foundation (No. 52161037) of China. The authors acknowledge particularly the HPC support from Dr. Miguel Dias Costa. The computational work for this article was performed on resources at the National Supercomputing Centre of Singapore (NSCC) and Centre for Advanced 2D Materials. This research was supported in part through computational resources of HPC facilities at HSE University.

**Data and code availability**
All data and codes are available at https://rolos.com/open/2d-materials-point-defects/ .

**Competing Interests**
The authors declare no competing financial or non-financial Interests.

**Authors contribution**
P.H., A.U. and K.S.N. conceived the research; P.H. and R.L. done most of the calculations, P.H., R.L., M.F., N.K. and A.R.A.-M. participated in data analyses, D.V.A., A.T. and A.H.C.N participated in the discussion. All authors contributed to the drafting of the work and approved the final version of the manuscript.